# Superconducting properties of the spin Hall candidate Ta$_3$Sb with eightfold degeneracy

R. Chapai,[1,*] A. Rydh,[2] M. P. Smylie,[1,3] D. Y. Chung,[1] H. Zheng,[1] A. E. Koshelev,[1] J. E. Pearson,[1] W.-K. Kwok,[1] J. F. Mitchell,[1] U. Welp[1,†]

[1]*Materials Science Division, Argonne National Laboratory, Lemont, IL 60439, USA*
[2]*Department of Physics, Stockholm University, Stockholm, Sweden*
[3]*Department of Physics and Astronomy, Hofstra University, Hempstead, NY 11549, USA*

We report the synthesis and characterization of phase pure Ta$_3$Sb, a material predicted to be topological with eightfold degenerate fermionic states [Science 353, aaf5037 (2016)] and to exhibit a large spin Hall effect [Sci. Adv. 5, eaav8575 (2019)]. We observe superconductivity in Ta$_3$Sb with $T_c \sim 0.67$ K in both electrical resistivity $\rho(T)$ and specific heat $C(T)$ measurements. Field dependent measurements yield the superconducting phase diagram with an upper critical field of $H_{c2}(0) \sim 0.95$ T, corresponding to a superconducting coherence length of $\xi \approx 18.6$ nm. The gap ratio deduced from specific heat anomaly, $2\Delta_0/k_B T_c$ is 3.46, a value close to the Bardeen-Cooper-Schrieffer (BCS) value of 3.53. From a detailed analysis of both the transport and thermodynamic data within the Ginsburg-Landau (GL) framework, a GL parameter of $\kappa \approx 90$ is obtained identifying Ta$_3$Sb as an extreme type-II superconductor. The observation of superconductivity in an eightfold degenerate fermionic compound with topological surface states and predicted large spin Hall conductance positions Ta$_3$Sb as an appealing platform to further explore exotic quantum states in multifold degenerate systems.

[*]*rchapai@anl.gov*; [†]*welp@anl.gov*

**Introduction**

In recent years, it has been realized that crystal symmetry, time-reversal symmetry and parity symmetry allow for the existence of novel quasiparticles in crystalline solids that are analogous to particles predicted in high-energy physics such as Dirac, Weyl, and Majorana fermions [1-3]. These exotic particles in crystalline solids are protected by band topology and crystal symmetry. With the growing understanding of topological properties in solids, particles which are even forbidden in high-energy physics due to Poincaré symmetry have also been predicted [4, 5]. Real materials realizations include triple point fermions in rock salt β-type GeTe [6] and WC-type MoP [7]; and sextuple point fermions in B20-type CoSi [8] and pyrite-type $PdSb_2$ [9, 10]. Recently, eightfold fermionic excitations have been predicted in several intermetallic compounds with the A15 structure including $Nb_3Bi$ and $Ta_3Sb$ [4, 11].

From a more practical viewpoint, topological materials are also potential platforms for quantum computing and spintronic applications [12]. In particular, nonmagnetic materials exhibiting large intrinsic spin Hall conductance (SHC) enabling efficient conversion of charge current to spin current are more desirable for spintronic applications [13, 14]. A recent theoretical study [11] indicates that non-magnetic A15 compounds display a large intrinsic SHC attributed to the large spin Berry curvature. Extraordinary values of -1400 $\hbar$/e (*S*/cm) and -2250 $\hbar$/e (S/cm) are anticipated for $Ta_3Sb$ and $W_3Ta$, respectively [11]. Moreover, band structure calculations indicate that $Ta_3Sb$ houses topological surface states (TSS) on the (001) surfaces connecting the conduction and valence bands along X-Γ-M in the Brillouin zone (BZ) [11,15]. These TSS derive from a nontrivial strong $Z_2$ topological invariant and exhibit nonhelical spin texture due to the reduced symmetry at the (001) surface [15]. In addition, $Ta_3Sb$ has been predicted to possess eightfold degenerate fermions stabilized by a non-symmorphic symmetry ($Pm\bar{3}n$) and is known to superconduct below 0.7 K [16]; however, the characteristic properties of the superconducting state such as critical fields, coherence length and superconducting gap are currently not known.

The expectation of nontrivial topology in $Ta_3Sb$ makes this material a potential candidate to explore topological superconductivity [17, 18]. Such a rare exotic state arises when the superconducting order parameter has non-trivial topological character such as a chiral *p*-wave order parameter proposed for $UTe_2$ [19, 20] or two-component nematic order envisioned for the doped topological insulators $M_xBi_2Se_3$ (M = Cu, Sr, Nb) [21]. Alternatively, materials with topologically non-trivial electronic band structures may display topological surface states with spin-polarized textures. A topological superconducting surface state can be induced into these surface states via proximity when the bulk of the material undergoes a transition into a conventional s-wave superconducting state. This scenario has been reported for various superconductors including the (010-surface of $MgB_2$ [22], the (001)-surface of $FeSe_xTe_{1-x}$ [23] and the (001) surface of non-centrosymmetric $PbTaSe_2$ [24]. However, experimental evidence for such exotic quantum states has yet to be obtained for $Ta_3Sb$.

In this work, we report the synthesis and characterization of high purity $Ta_3Sb$. Superconductivity is observed with $T_c$~0.67 K in both electrical resistivity $\rho(T)$ and specific heat $C(T)$ measurements confirming its bulk nature. Field dependent measurements of $\rho(T)$ and $C(T)$ yield the superconducting phase diagram, revealing an upper critical field of $\mu_0H_{c2}(0)$ ~0.95 T corresponding to a coherence length of $\xi \approx 18.6$ nm. In the vicinity of transition temperature, the $H_{c2}$(T) phase boundary displays an unusual downward curvature possibly caused by granular morphology of our samples. From the superconducting condensation energy obtained from specific heat, a superconducting gap of $\Delta_0$ ~ 0.1 meV is estimated yielding a gap ratio $2\Delta_0/k_BT_c$ = 3.46 that is very close to the Bardeen-Cooper-Schrieffer (BCS) weak-coupling value of 3.53. The step height ratio $\Delta C/\gamma T_c \approx 1.38$, is also close to the BCS expectation of 1.43 for a weak-coupling single-band isotropic *s*-wave superconductor [25]. From a detailed analysis of both the transport and thermodynamic data, a Ginzburg-Landau (GL) parameter of $\kappa \approx 90$ is estimated, identifying $Ta_3Sb$ as an extreme type-II superconductor. The bulk superconductivity we observed in a spin Hall candidate with TSS and multifold band crossing near the Fermi level ($E_F$) further stimulates studies on exploring topological properties of $Ta_3Sb$ and other sister compounds in the broad class of A15 family.

**Experimental details**

Polycrystalline $Ta_3Sb$ was prepared by thermal decomposition of $TaSb_2$ [16, 26]. High purity tantalum (99.98% Alfa Aesar) and antimony (99.95% Alfa Aesar) powders were mixed in a molar ratio of 1:2 and loaded into an alumina crucible, which was then sealed in a quartz tube under a vacuum of ~1 millitorr. The tube was heated to 800 °C at a rate of 100 °C/h in a box furnace, and held at 800 °C for 10 hrs. The temperature was then lowered to room temperature at a rate of 150°C/h. Powder x-ray diffraction (XRD) measurement performed in a PANalytical

X'pert Pro x-ray diffractometer with Cu K$_\alpha$ radiation confirmed the pure monoclinic phase (*C2/m*) of TaSb$_2$.

Thermal decomposition of TaSb$_2$ was accomplished in a vertical tube furnace with an opening at the top. Fine powder of TaSb$_2$ was loaded in an alumina crucible and sealed under vacuum in a quartz tube (~30 cm long, 12 mm OD, 9 mm ID). The lower end of the tube containing TaSb$_2$ was slowly heated to 1000°C at a rate of 50 °C/h. The other end of the tube was kept exposed to room temperature, thus creating the temperature gradient necessary for transporting and condensing volatile antimony. The furnace was maintained at 1000 °C for 14 days, followed by slow cooling to room temperature. The Antimony evolved from TaSb$_2$ was found deposited at the cold end, leaving behind Ta$_3$Sb at the hot end. XRD measurements confirmed the phase purity of both products Sb and Ta$_3$Sb. The pure phase Ta$_3$Sb thus obtained was further analyzed for the actual chemical composition through energy dispersive x-ray spectroscopy (EDS) measurement. Ta$_3$Sb in powder form was then pressed into small thin pellets and sintered at 900°C under vacuum for 24 hours for transport and specific heat measurements.

Electrical resistivity was measured using the standard four-probe method. Thin gold wires (50 μm) were attached to the sample using a conductive silver epoxy (Epotek H20E). The low-temperature transport measurements were performed in a Bluefors LD250 dilution cryostat using an ac-technique at 16.7 Hz while high-temperature transport data were acquired using a dc-technique in a Physical Properties Measurement System (DynaCool-PPMS, Quantum Design). Specific-heat measurements on microgram-sized samples were performed using a differential membrane-based nanocalorimeter operating in the fixed-phase ac steady-state mode. The calorimeter is composed of two cells, for the sample and the reference, respectively, each consisting of a thin-film cermet thermometer, an offset (direct current) heater, an ac heater and a thermalization layer, as described in Refs. [27, 28]. The sample was cut to a suitable size and mounted with Apiezon N grease onto the calorimeter. The measurements were performed in a Bluefors LD250 dilution refrigerator sitting at the base temperature with local temperature control obtained through the use of the offset heaters.

**Results and discussion**

Figure 1(a) shows the room temperature powder XRD pattern of single phase Ta$_3$Sb. The crystal structure was refined in the A15 structure type (*Pm$\bar{3}$n*, space group 223) using the Rietveld method on the XRD data ($R_{wp}$ = 4.56%, goodness-of-fit $\chi^2$ = 1.25), with *a* = 5.274(2) Å which is about 0.2% larger than that reported previously in Ref. [26]. In this structure, Sb atoms form a body centered cube, and the Ta atoms form one-dimensional chains in three orthogonal directions, with an interatomic spacing along the chains of $a/2$ [29]. A schematic crystal structure is depicted in the inset of Fig. 1 (a) with Sb atom marked red and Ta atom marked blue. The distribution of individual elements as well as the chemical composition in $Ta_3Sb$ was examined via EDS analysis. The EDS mapping of elemental composition shows uniform distribution of tantalum and antimony (lower right inset of Fig. 1(b)) throughout the sample. The EDS data presented in Fig. 1(b) gives the chemical composition ratio of Ta : Sb ~3.1(2) : 0.8(9) indicating a slight Sb deficiency which is consistent with the refinement of XRD data indicating Ta/Sb site mixing as described in the supplementary information [30, 31]. The possibility of site mixing of Ta on the Sb site is implied in the Ta-Sb phase diagram reported in Ref. [32], where the phase is found at $\approx$ 20 mol % Sb rather than at the nominal 25 mol %.

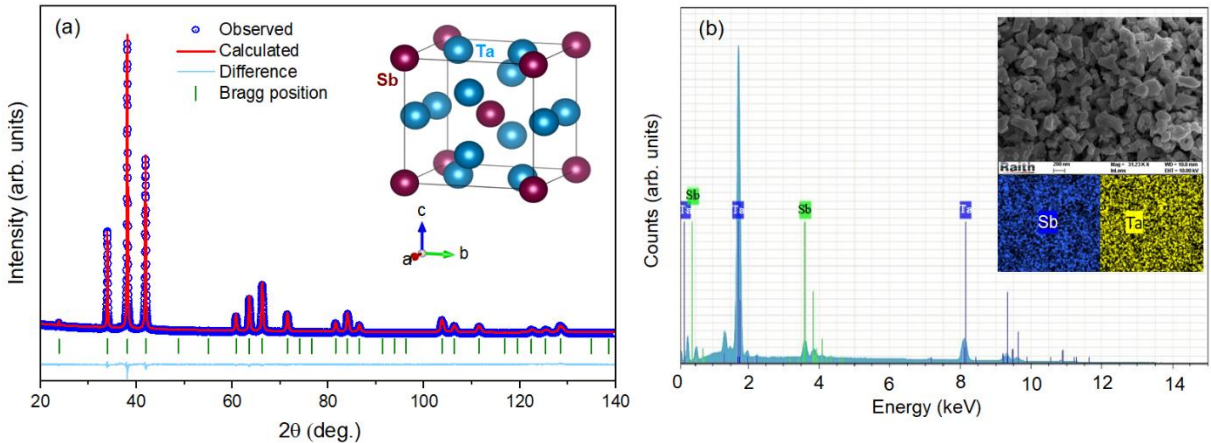

Figure 1. (a) Room temperature powder XRD pattern of $Ta_3Sb$ and Rietveld analysis profiles. Inset: Crystal structure of A15 type $Ta_3Sb$. (b) EDS spectrum of $Ta_3Sb$. Inset (upper right): SEM image of a typical press-pellet of $Ta_3Sb$ showing grain size. Inset (lower right): EDS mapping of elemental composition in $Ta_3Sb$.

Figure 2(a) shows the temperature dependent electrical resistivity between room temperature and 80 mK. The resistivity tends towards saturation at high temperatures at values above ~500 μΩcm and at low temperatures at a value of ~380 μΩcm resulting in a low residual resistivity ratio (RRR) of $\rho(300\ K)/\rho(2\ K) \approx 1.3$. The observed resistivity values exceed the Ioffe-Regel limit of 100-150 μΩcm seen in numerous A15 materials [33]. We attribute the high resistivity to a dominant scattering contribution due to grain boundaries in our compacted sample.

As seen in the scanning electron microscope (SEM) image (upper right inset of Fig. 1(b)), the sample is composed of grains with a characteristic size of ~60 nm, albeit with substantial variation in grain size and void pattern across the sample. Assuming that grain boundary scattering is weakly temperature dependent, then one can attribute the temperature dependent contribution to the resistivity seen in Fig. 2(a) as intrinsic to $Ta_3Sb$. Indeed, the change in resistivity with temperature is in good agreement with reported values for other A15s [33]. Notwithstanding its granular nature, the sample displays a sharp superconducting transition with $T_c$~0.67 K, a value close to that reported earlier in Ref. [16]. With the application of magnetic field, the transition weakly broadens while shifting to lower temperatures (Fig. 2(b)). To obtain the superconducting phase diagram (discussed later) down to the lowest temperature, we performed resistivity measurements as a function of magnetic field at various low temperatures below $T_c$ (Fig. 2(c)).

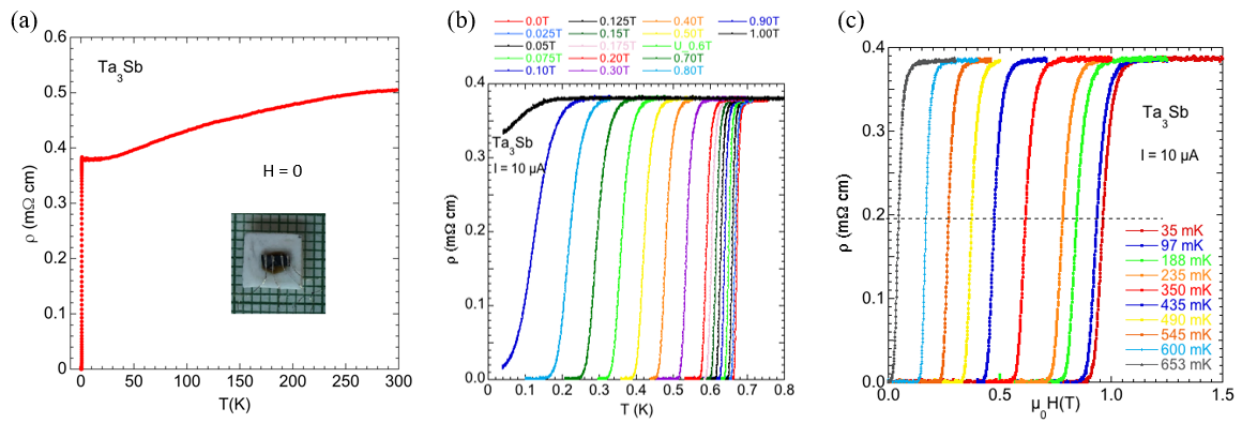

Figure 2. (a) Temperature dependence of electrical resistivity ρ (*T*) of $Ta_3Sb$ at zero field. Inset: a typical pressed pellet of $Ta_3Sb$ prepared for the resistivity measurement. (b) Temperature dependence of the resistivity through the superconducting transition in various applied magnetic fields as indicated. (c) Superconducting transition measured as a function of magnetic field at various temperatures. The dotted line indicates the definition of $T_c$ as the mid-point of the transitions.

To explore the emergence of possible superconducting surface states we mapped the resistive transitions for a wide range of applied current [30]. Conventional surface superconductivity arises in type-II superconductors due to boundary effects [34] and is revealed in the surface upper critical field $H_{c3}$ as well as in the temperature, field, and angular dependence of the critical current [35, 36]. Additional features may arise due to topological surface states (see above). However, as shown in Fig. S1 [30], the shape of the resistive transitions of our samples does not change with current (apart from a uniform shift at the highest currents due to self-heating) giving no indication for a superconducting surface state. Furthermore, characteristic differences in the temperature and field-angle dependence of the superconducting transition determined from thermodynamic and from transport measurements are expected [37, 38]. As described below, transport and specific heat measurements on our Ta$_3$Sb sample yield essentially identical phase boundaries suggesting that the investigation of possible surface states will require single-crystal samples with well-defined surfaces.

To obtain further insight into the superconducting properties of Ta$_3$Sb, we measured thermodynamic properties in both the high and low temperature regimes. Owing to the small sample size and its porosity, a direct measure of its volume/weight could not be obtained. Instead, we scale the measured heat capacity to the Dulong-Petit value which for Ta$_3$Sb is 100 J/(mol-f.u. K) yielding for our sample a mass of 1.5 10$^{-9}$ mol-f.u. corresponding to a volume of 6.57 10$^{-8}$ cm$^3$. Figure 3(a) displays the heat capacity $C(T)$ measured in the temperature range between 200 K and 150 mK. Superconductivity is observed in $C(T)$, confirming its bulk nature. The $T_c$ value obtained in $C(T)$ is consistent with that observed in $\rho(T)$. The normal state heat capacity in a range of temperatures above $T_c$ fits the conventional expression $C/T = \gamma + \beta T^2$ well (see bottom inset of Fig. 3(a)) where $\gamma$ is the normal-state electronic contribution and $\beta T^2$ is the lattice contribution to the specific heat. The fitting yields a Sommerfeld coefficient $\gamma$ = 4.3 mJ/(mol-f.u. K$^2$) and $\beta$ = 1 mJ/(mol-f.u. K$^4$). The $\gamma$ value is close to that of isostructural Ti$_3$Ir [39] while somewhat smaller than that of Nb$_3$Sn [29]. Nevertheless, this value of $\gamma$ = 4.3 mJ/(mol-f.u. K$^2$) identifies Ta$_3$Sb as a material with moderately high electronic density of states. Using the Debye relation $\beta$ = 12/5 $\pi^4 n_{fu} R/\theta_D^3$ [40], we obtain a Debye temperature of $\theta_D$ = 198 K. Here, $n_{fu}$ is the number of atoms per formula unit and $R$ is the molar gas constant. This value of $\theta_D$ is somewhat smaller than reported for instance for iron-based superconductors, 250 K- 290 K [41], or elemental Ta and Nb, 246 K and 276 K [42], respectively.

Figure 3(b) shows the temperature dependence of the difference of the specific heat in the superconducting ($C_s$) and normal state ($C_n$) in various applied fields plotted as ($C_s$-$C_n$)/$T$ vs $T$. We determine $T_c$ by approximating the measured data with an ideal step and an entropy-conserving construction as indicated by dotted lines in the figure, and find $T_c$ = 0.67 K. The resulting step height is $\Delta C/T_c \sim$ 6 mJ/(mol-f.u. K$^2$) corresponding to a ratio $\Delta C/\gamma T_c \sim$ 1.38 which is close to the BCS value of 1.43 for a weak-coupling single-band isotropic s-wave superconductor [25].

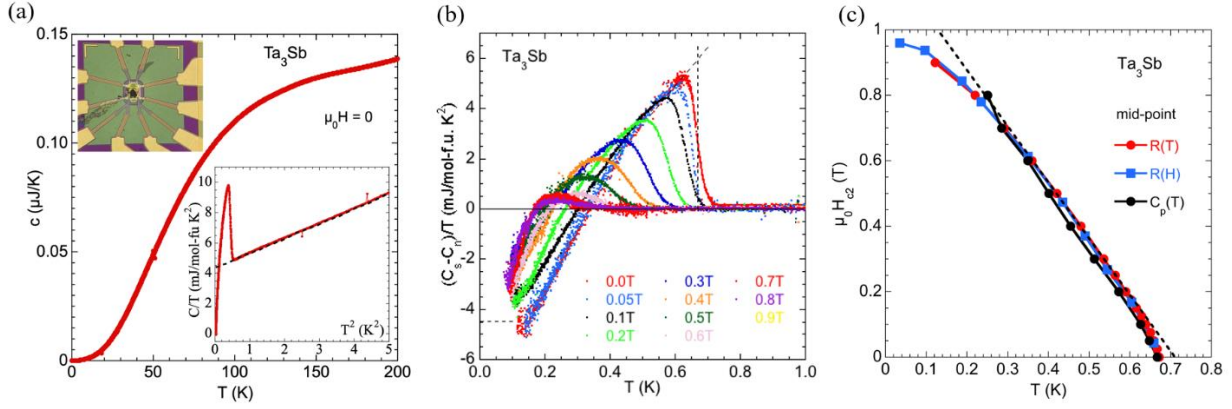

Figure 3. (a) Temperature dependence of the heat capacity *c(T)* at zero field. Top inset: A photograph of the sample mounted on the membrane nano-calorimeter. Bottom inset: specific heat *C/T* versus *T*$^2$, the dotted line indicates the determination of the Sommerfeld coefficient γ. (b) Temperature dependence of the difference of the superconducting and normal state specific heats in various applied fields. The horizontal dashed line represents the value of -γ. (c) The superconducting phase diagram of Ta$_3$Sb as determined from the mid-points in *R(T)*, *R(H)* and specific heat data. The downwards curvature of *H*$_{c2}$ near *T*$_c$ due to granularity appears in all three data sets. The dashed line indicates the estimate of the upper critical field slope.

The superconducting condensation energy at zero-temperature is given as an integral of the specific heat, $U(0) = \int_0^{>Tc}(C_s - C_n)dT$ yielding *U* (0) ~ 0.44 mJ/mol-f.u. As large experimental noise at temperatures below ~150 mK prevents a numerical integration, we approximated the data in this temperature range as (C$_s$-C$_n$) = -γT. Any possible error stemming from this approximation has a negligible contribution to the entire integral. This approximation, indicated as dashed line in Fig. 3(b), yields almost exact entropy conservation at the superconducting transition. The condensation energy defines the thermodynamic critical magnetic field *B*$_c$(0) via $U(0) = -B_c(0)^2/(2\mu_0)$ (μ$_0$ is the vacuum permeability) resulting in *B*$_c$(0) = -5 mT. Additional information on the thermodynamic critical field can be obtained from Rutger's relation $\frac{\Delta C}{T_c} = \frac{V}{\mu_0}\left(\frac{\partial B_c}{\partial T}\right)^2_{T_c}$ (*V* ≈ 43.8 cm$^3$ is the molar volume) from which a slope of $\partial B_c/\partial T|_{Tc}$ = -13 mT/K follows. We then estimate the reduced thermodynamic critical field $\frac{B_c(0)}{\left(\frac{\partial B_c}{\partial T}\right)_{T_c} T_c}$ = 0.56, while BCS theory for a weak-coupling single-band isotropic s-wave superconductor predicts a value of 0.576 [43]. The condensation energy can also be expressed in terms of microscopic parameters as *U*(0) = (*N*(0) Δ$_0$$^2$)/2, where *N*(0) is the electronic density of states at the Fermi level and Δ$_0$ is the zero-temperature superconducting gap. Using the relation between Sommerfeld coefficient and density of states, $\gamma = 2\pi^2 k_B{}^2 V N(0)(1+\lambda_{ep})/3$, we arrive at a gap value of $\Delta_0 = 0.1$ meV and a gap ratio of $2\Delta_0/k_B T_c = 3.46$, which is very close to the BCS value of 3.53. In this estimate, we have neglected a possible mass enhancement due to electron-phonon coupling $\lambda_{ep}$ since above values of the reduced thermodynamic critical field and of the normalized specific heat anomaly indicate weak coupling. In this case, $\gamma$ directly yields the density of states, $N(0) = 0.012$ 1/eV Å$^3$. In all, the analysis of the caloric data indicates that Ta$_3$Sb is a weak-coupling BCS superconductor. Further support for this assertion can be seen in the dimensionless quantity $-\frac{\mu_0 \gamma}{V}\left(\frac{T_c}{B_c(0)}\right)^2 = 0.168$ contained in BCS theory [43], for which our experimental value is 0.22. With increasing applied magnetic field, the specific heat anomaly shifts to lower temperatures and broadens, and its height is readily suppressed. Such behavior is commonly seen in field-dependent specific heat measurements on conventional superconductors [44] and is a consequence of entropy conservation. The observed height of the specific heat decreases linearly with field. This is the expected behavior of type-II superconductors since the electronic specific heat increases proportional to B, because in the vortex state the number of normal electrons increases with field approximately as $B/B_{c2}$, since the integral of $(C_s-C_n)/T$ over the total temperature range below $T_c$ has to be zero due to the entropy conservation, implying that this low-field increase has to be compensate by linear decrease of the specific heat jump at $T_{c2}(B)$.

From the field-dependent data shown in Fig. 2(b), Fig. 2(c), and Fig. 3(b), we construct the superconducting phase diagram of Ta$_3$Sb using the mid-point as criterion of $T_c$. All three data sets yield boundaries that closely track each other. We note that adopting a different criterion (onset, off-set, peak) will shift the phase boundary along the temperature axis but does not change its shape in a significant way. At low temperatures, the phase boundary flattens reminiscent of Werthamer-Helfand-Hohenberg (WHH) behavior [45], and approaches the value $B_{c2}(0) \sim 0.95$ T. At intermediate temperatures, the boundary is approximately linear with a slope of $\partial B_{c2}/\partial T \sim -1.7$ T/K extrapolating to a Ginzburg-Landau (GL) upper critical field of $B_{c2}^{GL} \sim 1.20$ T which corresponds to a GL coherence length of $\xi_{GL} \approx 16.5$ nm as obtained using the GL-relation $B_{c2} = \Phi_0/2\pi\xi_{GL}^2$ ($\Phi_0$ is the flux quantum). With the help of the relation $\partial B_{c2}/\partial T = 2^{1/2}\kappa\, \partial B_c/\partial T$ we estimate a GL parameter of $\kappa \sim 90$, a value far greater than $1/\sqrt{2}$ indicating that Ta$_3$Sb is an extreme type-II superconductor. Using $\kappa = \lambda_{GL}/\xi_{GL}$, we also can evaluate the GL value of the London penetration depth $\lambda_{GL} \approx 1.5$ μm.

The rather high value of the upper critical field slope suggests that paramagnetic limiting may play a role in determining the phase boundary. The paramagnetic limiting field ($H_P$) is given as $\mu_0 H_P = \Delta_0/\sqrt{2}\,\mu_B$ ($\mu_B$ is the Bohr magneton) which can be expressed as $\mu_0 H_P = 1.86\,T_c$ [46, 47] yielding $\mu_0 H_P \sim 1.3$ T. A signature of paramagnetic limiting is the pronounced flattening of the phase boundary at a value below the orbital critical field as given in WHH theory [45]. Even though the estimated paramagnetic limiting field and the measured upper critical field are comparable in size, our data do not reveal signatures of paramagnetic limiting. In fact, the measured upper critical field, 0.95 T, exceeds the expected orbital field [45] of 0.69 $B_{c2}^{GL} \sim 0.84$ T (dirty limit) and 0.72 $B_{c2}^{GL} \sim 0.86$ T (clean limit), respectively. Several scenarios may cause these observations. The expression $\mu_0 H_P = 1.86\,T_c$ underestimates the actual value of $H_p$ as it does not account for features including strong-coupling, anisotropy, spin-orbit scattering and multiple bands [48]. While our experimental data rule out strong-coupling effects, spin-orbit scattering is known to counteract paramagnetic limiting and increase the upper critical field towards the orbital value [49]. As Ta is a heavy element, spin-orbit coupling is strong, likely inducing significant spin-orbit scattering. Furthermore, the numerical coefficients of 0.69 (0.72) entering the WHH expression for the orbital critical field are derived for a spherical or ellipsoidal Fermi surface. We note that the value for clean limit is not universal and depends on the shape of Fermi surface. For instance, for a Fermi surface in the shape of a warped cylinder the coefficient approaches 1.18 at the Lifshitz neck-disrupting transition [50].

At temperatures above ~0.6 K, the phase boundary displays a clear downward curvature. The fact that this downwards curvature is also seen in the specific heat data indicates that it is a thermodynamic feature. A likely cause is associated with confinement effects arising from the granular nature of our sample. Several parameters such as grain size, Josephson coupling energy between grains and Coulomb charging energy, among others, determine the properties of granular superconductors [51, 52]. Because the sample is metallic in the normal state, charging effects are expected to be negligible. Furthermore, the occurrence of sharp resistive transitions even in high magnetic fields and the absence of resistive tails at high current (see supplementary information [30]) indicate that the grains are strongly coupled. Therefore, at low temperatures, when the coherence length is smaller than the typical grain size and/or the void spacing, the material behaves as a three-dimensional superconductor with a linear $B_{c2}$-curve (in GL theory). With increasing temperature, the coherence length $\xi_{GL}$ grows. However, the divergence associated with a 2$^{nd}$ order phase transition is cut-off at a typical length scale corresponding to the grain size and/or void spacing. At this cross-over point, the $B_{c2}$-curve acquires a $(T_c-T)^{1/2}$ -variation. A cross-over temperature of ~0.6 K would correspond to a length scale ≈ 45 nm comparable to the smallest feature sizes seen in the SEM images (inset of Fig. 1(b)). While the phase boundary displays downward curvature, it is not as sharp as a square root variation. We believe this is likely due to a (broad) distribution of grain sizes. Extrapolating the linear section of $B_{c2}(T)$ to high temperature, one would find a $T_c$ ~ 0.71 K (see the dashed line in Fig. 3(c)), which, in the scenario of confinement, would be the $T_c$ measured on a uniform single crystal sample. Previous work on superconducting particles has shown that the specific heat anomaly at the transition is suppressed and shifted to lower temperatures with decreasing particle size and decreasing inter-particle coupling [53-55]. Here, we use the $T_c$-suppression as a measure of the degree to which above analysis of the specific heat is affected by confinement effects and conclude that an uncertainty of ~10% could arise.

In summary, we have systematically characterized the superconducting properties of phase pure Ta$_3$Sb. Based on measurement of transport and thermodynamic quantities, several parameters characterizing the superconducting state have been evaluated within GL theory. We find a superconducting coherence length of $\xi_{GL}$ ~18.6 nm and a superconducting gap of 0.1 meV. Our experimental observations suggest that Ta$_3$Sb is a weakly coupled BCS superconductor with extreme type II nature. Whether or how the predicted topological band structure contributes to superconductivity in Ta$_3$Sb is not apparent from our current results. We envision that Ta$_3$Sb is in the same group of materials as MgB$_2$ and FeSe$_x$Te$_{1-x}$ in which *s*-wave superconductivity in the bulk induces topological surface superconductivity [15] which can be revealed in spectroscopic techniques such as scanning tunneling microscopy (STM) and angle resolved photoemission spectroscopy (ARPES). These measurements as well as the exploration of the predicted giant intrinsic spin Hall effect would require single crystalline samples. Besides, it is worth noting that the non-stoichiometry confirmed by both EDS and Rietveld analysis will affect at the very least the electron-count in band structure calculations ($E_F$ should be higher than that calculated for stoichiometric Ta$_3$Sb due to a larger number of electrons in a Ta-rich phase), potentially opening a chemical route to tune the band structure. We note that Nb$_3$Sn forms in the A15 phase for Sn-concentrations ~ 0.18 $\leq \beta \leq$ ~ 0.26 with $\beta$ defined via Nb$_{1-\beta}$Sn$_\beta$. Over this compositional range, $T_c$ increases from ~5 K to 18 K, and superconductivity evolves from weak-coupling to strong-coupling [56].

**Acknowledgements**

This work was supported by the U. S. Department of Energy, Office of Science, Basic Energy Sciences, Materials Sciences and Engineering Division. The EDS measurement performed at the Center for Nanoscale Materials, a U. S. Department of Energy, Office of Science User Facility, was supported by the U. S. DOE, Office of Basic Energy Sciences, under Contract No. DE-AC02-06CH11357. Heat capacity measurement were supported by the Swedish Research Council, D. Nr. 2021-04360.

**References**

1. Z. K. Liu, B. Zhou, Y. Zhang, Z. J. Wang, H. M. Weng, D. Prabhakaran, S.-K. Mo, Z. X. Shen, Z. Fang, X. Dai *et al*., Discovery of a Three-Dimensional Topological Dirac Semimetal, Na$_3$Bi, *Science* **343,** 864 (2014).

2. V. Mourik, K. Zuo, S. M. Frolov, S. R. Plissard, E. P. A. M. Bakkers, and L. P. Kouwenhoven, Signatures of Majorana Fermions in Hybrid Superconductor-Semiconductor Nanowire Devices, *Science* **336**, 1003 (2012).

3. B. Q. Lv, H. M. Weng, B. B. Fu, X. P. Wang, H. Miao, J. Ma, P. Richard, X. C. Huang, L. X. Zhao, G. F. Chen *et al*., Experimental Discovery of Weyl Semimetal TaAs, *Phys. Rev. X* **5**, 031013 (2015).

4. B. Bradlyn, J. Cano, Z. Wang, M. G. Vergniory, C. Felser, R. J. Cava, and B. A. Bernevig, Beyond Dirac and Weyl fermions: Unconventional quasiparticles in conventional crystals, *Science* **353**, aaf5037 (2016).

5. Q. Liu and A. Zunger, Predicted Realization of Cubic Dirac Fermion in Quasi-One-Dimensional Transition-Metal Monochalcogenides, *Phys. Rev. X.* **7**, 021019 (2017).

6. J. Krempaský, L. Nicolaï, M. Gmitra, H. Chen, M. Fanciulli *et al*., Triple-Point Fermions in Ferroelectric GeTe, *Phys. Rev. Lett*. **126**, 206403 (2021).

7. N. Kumar, Y. Sun, M. Nicklas, S. J. Watzman, O. Young *et al*., Extremely high conductivity observed in the triple point topological metal MoP, *Nat. Commun*. **10**, 2475 (2019).

8. B. Xu, Z. Fang, M.-A. S.-Martinez, J. W. F. Venderbos, Z. Ni *et al*., Optical signatures of multifold fermions in the chiral topological semimetal CoSi, *Proc. Natl. Acad. Sci*. **117**, 27104 (2020).

9. R. Chapai, Y. Jia, W. A. Shelton, R. Nepal, M. Saghayezhian, Fermions and bosons in nonsymmorphic $PdSb_2$ with sixfold degeneracy, *Phys. Rev. B* **99**, 161110(R) (2019).

10. X. Yáng, T. A. Cochran, R. Chapai, D. Tristant, J-X. Yin, Observation of sixfold degenerate fermions in $PdSb_2$, *Phys. Rev. B* **101**, 201105(R) (2020).

11. E. Derunova, Y. Sun, C. Felser, S. S. P. Parkin, B. Yan, and M. N. Ali, Giant intrinsic spin Hall effect in $W_3Ta$ and other A15 superconductors, *Sci. Adv*. **5**, eaav8575 (2019).

12. A. Fert, Nobel Lecture: Origin, development, and future of spintronics. *Rev. Mod. Phys*. **80**, 1517 (2008).

13. H. Ohno, M. D. Stiles, B. Dieny, Spintronics. *Proc. IEEE Inst. Electr. Electron. Eng*. **104**, 1782 (2016).

14. Y. Huai, Spin-transfer torque MRAM (STT-MRAM): Challenges and prospects. *AAPPS Bull.* **18**, 33 (2008).

15. M. Kim, C.-Z. Wang, and K.-M. Ho, Topological states in A15 superconductors, *Phys. Rev. B* **99**, 224506 (2019).

16. H. L. Luo, E. Vielhaber, and E. Corenzwit, Occurrence of A15 Phases, *Z. Physik* **230**, 443 (1970).

17. M. Sato and Y. Ando, Topological superconductors: a review, *Rep. Prog. Phys*. **80**, 076501 (2017)].

18. K. Flensberg, F. V. Oppen and A. Stern, Engineered platforms for topological superconductivity and Majorana zero modes, *Nat. Rev. Mater.* 6, 944 (2021.

19. S. Ran, C. Eckberg, Q.-P. Ding, Y. Furukawa, T. Metz *et al*., Nearly ferromagnetic spin-triplet superconductivity, *Science* **365**, 684 (2019).

20. L. Jiao, S. Howard, S. Ran, Z. Wang, J. Olivares Rodriguez *et al*., Chiral superconductivity in heavy-fermion metal $UTe_2$, *Nature* **579**, 523 (2020).

21. S. Yonezawa, Review: Nematic Superconductivity in Doped $Bi_2Se_3$ Topological Superconductors, *Condens. Matter* **4**, 2 (2019).
22. X. Zhou, K. N. Gordon, K.-H. Jin, H. Li, D. Narayan *et al.*, Observation of topological surface states in the high-temperature superconductor $MgB_2$, *Phys. Rev.* B **100**, 184511 (2019).
23. P. Zhang, K. Yaji, T. Hashimoto, Y. Ota, T. Kondo *et al.*, Observation of topological superconductivity on the surface of an iron-based superconductor, *Science* **360**, 182 (2018).
24. T.-R. Chang, P.-J. Chen, G. Bian, S.-M. Huang, H. Zheng *et al.*, Topological Dirac surface states and superconducting pairing correlations in $PbTaSe_2$, *Phys. Rev*. B 93, 245130 (2016).
25. M. Tinkham, *Introduction to superconductivity* (Dover Publication, New York 2004).
26. S. Furuseth, K. Selte, and A. Kjekshus, On the Arsenides and Antimonides of Tantalum, *Acta Chem. Scand*. **19**, 106 (1965).
27. K. Willa, Z. Diao, D. Campanini, U. Welp, R. Divan, M. Hudl, Z. Islam, W.-K. Kwok, and A. Rydh, Nanocalorimeter platform for in situ specific heat measurements and x-ray diffraction at low temperature, *Rev. Sci. Instrum.* **88**, 125108 (2017).
28. S. Tagliati, V. M. Krasnov, A. Rydh, Differential membrane-based nanocalorimeter for high-resolution measurements of low-temperature specific heat. *Rev. Sci. Instrum.***83**, 055107 (2012).
29. G.R Stewart, *Phys. C* **514**, 28 (2015), Superconductivity in the A15 structure.
30. Supplemental information includes the Rietveld refinement of the XRD data and heating effect of electrical current (if any) in the sample during electrical resistivity measurement (which includes Ref. [31, 32]).
31. R. W. Cheary, A. Coelho, A fundamental parameters approach to X-ray line-profile fitting. *J Appl Cryst.* **25**, 109 (1992).
32. F. Failamani, P. Broz, D. Maccio, S. Puchegger, H. Muller *et. al.,* Constitution of the systems {V,Nb,Ta}-Sb and physical properties of di-antimonides {V,Nb,Ta}$Sb_2$, *Intermetallics* **65**, 94 (2015).
33. N. E. Hussey, K. Takenaka, H. Takagi, Universality of the Mott-Ioffe-Regel limit in metals, *Philos. Mag*. **84**, 2847 (2004).
34. D. Saint-James and P. G. Gennes, Onset of superconductivity in decreasing fields, *Phys. Lett.* **7**, 306 (1963).

35. C. F. Hempstead and Y. B. Kim, Resistive transitions and surface effects in type-II superconductors, *Phys. Rev. Lett.* **12**, 145 (1964).

36. H. R. Hart and P. S. Swartz, Studies of surface transport currents in type-II superconductors; A surface-flux-pinning model, *Phys. Rev.* **156**, 403 (1967).

37. Y. J. Sato, F. Honda, Y. Shimizu, A. Nakamura, Y. Homma *et al.*, Anisotropy of upper critical field and surface superconducting state in the intermediate-valence superconductor CeIr$_3$, *Phys. Rev.* B **102**, 174503 (2020).

38. A. Rydh, U. Welp, J. M. Hiller, A. E. Koshelev, W. K. Kwok *et al.*, Surface contribution to the superconducting properties of MgB$_2$ single crystals, *Phys. Rev.* B **68**, 172502 (2003).

39. M. Mandal, K. P. Sajilesh, R. R. Chowdhury, D. Singh, P.K. Biswas, A.D. Hillier, and R. P. Singh, Superconducting ground state of the topological superconducting candidates Ti$_3$X (X= Ir, Sb), *Phys. Rev.* B **103**, 054501 (2021).

40. C. Kittel. *Introduction to Solid State Physics* (Wiley, New York, 2006).

41. S. L. Bud'ko, T. Kong, W. R. Meier, X. Ma, and P. C. Canfield, $^{57}$Fe Mössbauer study of stoichiometric iron-based superconductor CaKFe$_4$As$_4$: a comparison to KFe$_2$As$_2$ and CaFe$_2$As$_2$, *Philos. Mag.* **97**, 2689 (2017).

42. G. R. Stewart, Measurement of low-temperature specific heat, *Rev. Sci. Instrum.* **54**, 1 (1983).

43. J. P. Carbotte, Properties of boson-exchange superconductors, *Rev. Mod. Phys.* **62**, 1027 (1990).

44. A. Mirmelstein, A. Junod, E. Walker, B. Revaz, J. Y. Genoud, G. Triscone, Mixed-State Specific Heat of the Type-II Superconductor Nb$_{0.77}$Zr$_{0.23}$ in Magnetic Fields up to B$_{c2}$, *J. Supercond.* **10**, 527 (1997).

45. N. R. Werthamer, E. Helfand, and P. C. Hohenberg, Temperature and Purity Dependence of the Superconducting Critical Field, $H_{c2}$. III. Electron Spin and Spin-Orbit Effects, *Phys. Rev.* **147**, 295 (1966).

46. A. M. Clogston, Upper Limit for the Critical Field in Hard Superconductors, *Phys. Rev. Lett.* **9**, 266 (1962).

47. B. S. Chandrasekhar, A note on the maximum critical field of high-field superconductors, *J. Appl. Phys. Lett.* **1**, 7 (1962).

48. G. Fuchs, S. -L Drechsler, N. Kozlova, M. Bartkowiak, J. E. H.-Borrero *et al*., Orbital and spin effects for the upper critical field in As-deficient disordered Fe pnictide superconductors, *New J. of Phys.* **11**, 075007 (2009).

49. Ø. Fischer and M. Peter, in "Magnetism": A Treatise on Modern Theory and Materials, edited by H. Suhl (Academic Press, New York, 1973).

50. A. Bianconi and T. Jarlborg, Lifshitz transitions and zero-point lattice fluctuations in sulfur hydride showing near room temperature superconductivity. *Nov. Supercond. Mater.* **1**, 37 (2015).

51. G. Deutscher, Nanostructured Superconductors, in "Superconductivity" (editors K. H. Bennemann and J. B. Ketterson, Springer Verlag, Berlin 2008).

52. S. Bose, P. Ayyub, A review of finite size effects in quasi-zero-dimensional superconductors, *Rep. Prog. Phys.* **77**, 116503 (2014).

53. T. Worthington, P. Lindenfeld, G. Deutscher, Heat-Capacity Measurements of the Critical Coupling between Aluminum Grains, *Phys. Rev. Lett*. **41**, 316 (1978).

54. R. L. Filler, P. Lindenfeld, T. Worthington, G. Deutscher, Heat-capacity of granular aluminum, *Phys. Rev. B* **21**, 5031 (1980).

55. G. Deutscher, O. Entin-Wohlman, S. Fishman, and Y. Shapira, Percolation description of granular superconductors, *Phys. Rev. B* **21**, 5041 (1980).

56. A. Godeke, Performance Boundaries in $Nb_3Sn$ Superconductors, Ph.D. thesis, University of Twente, Enschede, The Netherlands, ISBN 90-365-2224-2 (2005), and references therein.